# Liquid Surface X-ray Studies of Gold Nanoparticle-Phospholipid Films at the Air/Water Interface

*Siheng Sean You[1], Charles T. R. Heffern[1,2], Yeling Dai[3], Mati Meron[4], J. Michael Henderson[1,2], Wei Bu[4], Wenyi Xie[1,5], Ka Yee C. Lee[1,2,6]\*, Binhua Lin[1,4]\**

[1] James Franck Institute, The University of Chicago, Illinois 60637, USA.

[2] Department of Chemistry, The University of Chicago, Illinois 60637, USA.

[3] Departament of Physics, The University of California San Diego, La Jolla, California 92093, USA.

[4] Center for Advanced Radiation Sources, The University of Chicago, Illinois 60637, USA.

[5] Department of Chemical Engineering, City College of New York, New York City, New York 10031, USA.

[6] Institute for Biophysical Dynamics, The University of Chicago, Illinois 60637, USA.

# Abstract

Amphiphilic phospholipids and nanoparticles functionalized with hydrophobic capping ligands have previously been extensively investigated for their capacity to self-assemble into Langmuir monolayers at the air/water interface. However, understanding of composite films consisting of both nanoparticles and phospholipids, and by extension, the complex interactions arising between nanomaterials and biological membranes, remains limited. In this work, dodecanethiol-capped gold nanoparticles (Au-NPs) with an average core diameter of 6 nm were incorporated into 1,2-dipalmitoyl-*sn*-glycero-3-phosphocholine (DPPC) monolayers in area ratios ranging from 0.1 to 20% area coverage at a surface pressure of 30 mN/m. High resolution liquid surface X-ray scattering studies revealed a phase separation of the DPPC and Au-NP components of the composite film, as confirmed with atomic force microscopy after the film was transferred to a substrate. At low Au-NP content, the structural organization of the phase-separated film is best described as a DPPC film containing isolated islands of Au-NPs. However, increasing the Au-NP content beyond 5% area coverage transforms the structural organization of the composite film to a long-range interconnected network of Au-NP strands surrounding small seas of DPPC, where the density of the Au-NP network increases with increasing Au-NP content. The observed phase separation and structural organization of the phospholipid and nanoparticle components in these Langmuir monolayers are useful for understanding interactions of nanoparticles with biological membranes.

# Introduction

Inorganic nanoparticles have been explored for numerous applications in the biological sciences ranging from drug delivery and transfection vehicles[1-6], to fluorescent labelling and imaging contrast agents[1, 5, 7-11]. Increasing use of nanoparticles in biological applications motivates exploration of the physical interactions between nanoparticles and biological systems, in particular the cell membrane. Nanoparticles have been found to interface with, adhere to, and penetrate phospholipid membranes[12-15]. Understanding of the effect of nanoparticles on the structure and ordering of lipid membranes is important to assessing the viability of their biological applications and potential cytotoxicity.

The air/water interface is an integral system for the study of the structure and organization of phospholipid films. Previous studies have examined the self-assembly and morphology of nanoparticle and phospholipid composite monolayers at the air/water interface using Brewster angle microscopy, neutron scattering and surface pressure measurements *in situ*[16-17] as well as atomic force microscopy (AFM)[18-19] and transmission electron microscopy (TEM)[18, 20] following transfer of the monolayer to an appropriate imaging substrate. These studies show that nanoparticles with hydrophobic capping ligands tend to phase separate in the lipid films as the nanoparticles become excluded from the condensed lipid domains, and that integration of nanoparticles into the lipid film tends to reduce the condensed lipid domain size.

Our understanding of lipid/nanoparticle composite films can be enhanced through the use of liquid surface X-ray scattering (LSXS) techniques, which can be performed *in situ* and provide information on the ordering and structure of both lipid and nanoparticle constituents of the film at sub-nanometer length scales.[21-22] LSXS has been used extensively to study the physical properties, structure and ordering of Langmuir monolayers of phosophlipids,[23-28] and, more recently, self-assembled pure nanoparticle films at the air/water interface.[29-35]

In this paper, we report the study of a mixed phospholipid/ nanoparticle system comprising gold nanoparticles functionalized with a surface ligand of dodecanethiol (Au-NPs) and 1,2-dipalmitoyl-*sn*-glycero-3-phosphocholine (DPPC), with Au-NP content ranging from 0.1 to 20% area coverage using LSXS and AFM. These materials were chosen as Au-NPs have been used in applications such as localized heating to induce apoptosis, drug delivery, and imaging indicators[1-3, 15, 36-39] and DPPC is a well-studied phospholipid, typically used to model biological lipid films ranging from cellular membranes to lung surfactant.[40-42]

While our Langmuir surface pressure/area isotherm demonstrated that the response of composite DPPC/Au-NP films upon uniaxial compression is dominated by DPPC, X-ray reflectivity (XR) and grazing incidence X-ray diffraction (GIXD) measurements showed that phase separation of the two monolayer components appear even at an Au-NP content of 0.1 % area coverage. Increasing Au-NP content increase size of the Au-NP domains, which in turns influences the packing patterns of the phospholipids, as seen through the change in the tilt of their tail groups. These observations of the DPPC/Au-NP composite film was corroborated with AFM imaging, which showed that the Au-NP formed phase separated, fractal-like structures which remained hexagonally packed. Through the use of LSXS techniques, we demonstrate that integration of Au-NP into a lipid monolayer at a sufficiently high content can affect the packing structure of the lipid molecules, despite the phase separation between the components of the composite film.

**Experimental**

**Materials**

DPPC was purchased in powder form from Avanti Polar Lipids (Alabaster, AL; chemical purity >99%), dissolved into HPLC grade chloroform (Fisher Scientific, Pittsburgh, PA) at a concentration of 1 mg/mL, and stored at -20°C. Gold nanoparticles with a core diameter of ~6 nm and functionalized with a surface ligand of dodecanethiol (Au-NP), which has a length of ~1.7 nm when fully extended, were purchased from Ocean Nanotech (San Diego, CA), suspended in HPLC grade chloroform to a concentration of 1 mg/mL, and stored at room temperature. For all experiments, ultrapure water (resistivity ≥ 18.2 MΩ-cm) obtained from a Milli-Q UV Plus system (Millipore, Bedford, MA) was used as the subphase. DPPC/Au-NP mixtures of these stock solutions were prepared immediately prior to experiments.

**Area Coverage Calculations**

The area coverage ratio between the DPPC and Au-NP was estimated as follows: Each DPPC lipid molecule was assumed to have an area of 47.3 Å$^2$ at 30 mN/m based on pure DPPC GIXD measurements shown in previous works;[23] DPPC area coverage was then estimated by multiplying this area by the number of molecules of DPPC, determined from the mass of DPPC deposited and its molecular weight. The nanoparticle area coverage was estimated in two steps: first the number of nanoparticles deposited was calculated; the area coverage for a single nanoparticle was then calculated and multiplied by the number of nanoparticles deposited. The mass of one nanoparticle ($M_{np}$), $M_{np} = \frac{4}{3}\pi r^3 \rho$, where $r$ is the nanoparticle core radius and $\rho$ is the density of gold, assuming a spherical and neglecting the ligand mass known to contribute <10% of total NP mass from previous studies.[43-44] The area coverage of a single nanoparticle is calculated assuming hexagonal packing of the nanoparticles: each nanoparticle occupies an area of $A_{np} = 2\sqrt{3}l^2$ where $l$ is the length of the apothem of the hexagon which inscribes the nanoparticle, the sum of ½ the core (6 nm) and ligand lengths (1.7 nm), assuming full interpenetration of the nanoparticle ligands which is consistent with the Au-NP spacing measured using GIXD. This yields an $A_{np}$ of ca. 205 nm$^2$. Table 1 provides the estimated area coverage ratio, mass ratio and mole ratio of the DPPC/Au-NP mixtures used in this study.

| Mass Ratio | Mole Ratio | DPPC/Au Calculated Area Ratio |
|---|---|---|
| 1:0.015 | 1:0.0000086 | 99.9:0.1 |
| 1:0.153 | 1:0.000087 | 99:1 |
| 1:0.39 | 1:0.00022 | 97.5:2.5 |
| 1:0.8 | 1:0.00045 | 95:5 |
| 1:1.23 | 1:0.0007 | 92.5:7.5 |
| 1:1.68 | 1:0.00096 | 90:10 |
| 1:3.79 | 1:0.0022 | 80:20 |

**Table 1.** Estimated surface area ratios of DPPC/Au-NP used in this work at 30 mN/m with corresponding mass and mole ratios.

**Surface Pressure-Area Isotherms**

Surface pressure-area isotherms were obtained with a custom-built, Teflon, Langmuir trough of dimensions 8.9 cm × 40 cm equipped with a Wilhelmy balance and described in greater detail elsewhere.[29, 31-32, 45] Surface pressure measurements were obtained by monitoring the difference in surface tension between a clean air/water interface and the surface tension of the interface with the Langmuir monolayer present. In a typical experiment, a solution of DPPC and Au-NP dissolved in chloroform to a concentration of ~0.5 mg/mL was spread at the air/water interface to a low surface density, leaving an initial surface pressure of 0 mN/m. Deposition and compression of the film was separated by 30 minutes to allow time for complete solvent evaporation and film relaxation. Compression of the film was achieved using an isoperimetric Teflon barrier moving at a rate of 10 cm$^2$/min, which translates to a rate of reduction of the mean molecular area between 0.05 and 0.06 Å$^2$/s, depending on the amount of material deposited at the interface. All experiments were performed at ambient temperature, 23°C.

**AFM Sample Preparation and Imaging**

Samples were prepared for AFM by compressing the film to a surface pressure of 30 mN/m and holding the film at this pressure through a surface pressure feedback loop for a minimum of 30 minutes to allow for film relaxation. Following that, the barrier was stopped and the sample transferred onto a high-grade mica substrate (Ted Pella Inc., Redding, CA) using the Langmuir-Schafer deposition method.[46-47] The sample was dried in air for a minimum of 1 hour before being imaged with a MultiMode Nanoscope IIIA Scanning Probe Microscope (Digital Instruments/Bruker, Santa Barbara, CA) using tapping mode with a Bruker OTESPA series probe.

**X-ray Scattering Measurements**

All liquid surface X-ray scattering measurements were conducted at the Advanced Photon Source (Argonne, IL) in ChemMatCARS, station 15-ID-C, using monochromatic X-rays with a wavelength of 1.24 Å (10k eV) and a beam footprint of ~100 mm$^2$ on the sample. Measurements were performed on the same custom-built, Teflon, Langmuir trough used for the surface pressure-area isotherms on films held at a constant surface pressure of 30 mN/m via a feedback loop. Before any X-ray data were collected, at least 30 minutes were allowed after a sample reached the target surface pressure to flush the trough enclosure with helium to reduce the scattering background and to ensure film relaxation. Both XR and GIXD measurements were performed on all films. The experimental geometry has been illustrated and described in detail previously.[25-26]

XR and GIXD were performed to obtain in-plane as well as out-of-plane electron density distribution information, respectively. XR data were analyzed using the Parratt method.[29, 31, 48-49] The electron density $\rho(z)$ was laterally averaged over the footprint of the X-ray beam, and modeled by a sum of error functions $\text{erf}(z) = 2/\sqrt{\pi} \int_0^z e^{-t^2} dt$ of the form:[21]

$$\rho(z) = \frac{1}{2} \sum_{i=0}^{N-1} \text{erf}\left(\frac{z - z_i}{\sqrt{2}\sigma}\right)(\rho_i - \rho_{i+1}) + \frac{\rho_0 - \rho_N}{2}$$

where $\sigma$ is the surface roughness from capillary wave theory, $N$ is the number of internal interfaces, $z_i$ is the position of the $i$th interface, $\rho_i$ is the electron density of the $i$th interface and $\rho_0$ is the electron density of the aqueous subphase.

The GIXD intensity resulting from a powder of 2-D crystallites may be represented by its projection onto the $q_{xy}$ axis to yield Bragg peaks or onto the $q_z$ axis to yield Bragg rods, where $q_{xy}$ and $q_z$ are the horizontal and vertical components of the scattering vector $q$, respectively.[25, 50] From the diffraction signal of the DPPC component of the composite films, the intensity distribution of the Bragg rods and peaks were analyzed to determine the $d$ spacing of the DPPC ($d = 2\pi / q_{xy}$), the correlation length of the crystalline domains, and the tilt angle, $\theta$, for the nearest neighbor tilt of the hydrocarbon tails.[23, 25, 50] Analysis of the Au-NP diffraction signal shows the presence of multiple Bragg scattering peaks. The center-to-center spacing of the Au-NP was obtained by performing a least squares fit to the first diffraction peak, and converting reciprocal space distance into real space.[32-33]

## Results and Discussions

In order to evaluate the mechanical properties of the composite monolayer, Langmuir isotherms of the composite films were compared to those of the pure components. The surface area–surface pressure isotherm of DPPC has been well studied, and has several distinct phases.[51-56] At low surface density, the DPPC remains in coexistence between the gas and the liquid-expanded (LE) phases; upon compression, it enters first the LE phase followed by a transition into a two-phase coexistence, indicated by a plateau in the surface pressure-surface area curve, to the condensed (C) phase, and finally collapses into the third dimension through cracking at high surface density and pressures (> 70 mN/m). In contrast, films of Au-NPs collapse into the third dimension at lower surface pressures (~30 mN/m).[30, 32-33, 45] This collapse occurs via a transition from a monolayer to a multilayer, as observed using optical microscopy and inferred by a decrease in the slope of the surface pressure-surface area isotherm.

For all mixed systems studied in this work, the isotherms retain the qualitative shape of the DPPC isotherm, including its distinct phases (Figure 1). The impact of the Au-NPs on the mixed monolayer isotherms is minimal at low pressures, excluding the change in slope around the LE-C coexistence plateau (Figure 1) and the increase in the liftoff area due to surface area occupied by the Au-NP. This indicates that the Au-NPs do not contribute significantly to the intermolecular interactions until high Au-NP surface density. The steepness of the surface pressure-surface area curve through the LE-C coexistence plateau increases with increasing area coverage of Au-NPs (0.1 - 20%), possibly due to the stress response of the Au-NPs phase separated islands (Figure 1); While the DPPC in the mixed film undergoes a phase transition at the LE-C plateau, there is no corresponding phase transition for the Au-NPs, resulting in the mixed films displaying a stress response to compression in what was previously a stress-free phase transition. Furthermore, the pressure response in the LC phase of the higher Au-NP surface density films (10 % and 20 % Au-NP), show a kink around 35-40 mN/m. This feature maybe result from the transition of the Au-NP monolayer islands into Au-NP multilayers, as the monolayer to multilayer transition for pure NP films has been shown to appear at this pressure range.[29-30, 32, 44-45, 57]

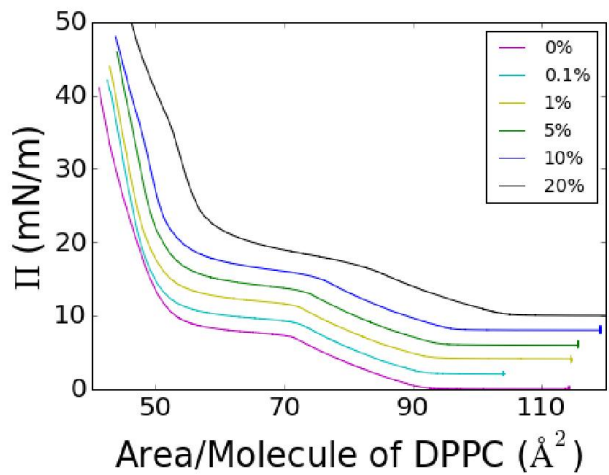

**Figure 1**: $\pi - A$ isotherms of DPPC/Au-NP composite films at 23°C with estimated Au-NP area coverages 0% (yellow), 0.1% (pink), 1% (teal), 5% (red), 10% (green), and 20% (blue). Each isotherm is offset vertically by 2 mN/m for viewing clarity. As the Au-NP content in the composite film is increased, the $\pi - A$ isotherms retain the overall shape of the pure DPPC isotherm, with the exception of the slope of the isotherm in the DPPC phase transition plateau at ~ 9 mN/m. The increasing liftoff area with increasing Au-NP content reflects the additional area occupied by the Au-NPs.

The phase separation of the Au-NP and DPPC components of the mixed monolayers observed by previous studies was first confirmed with high resolution AFM imaging (Figure 2). Two distinct phases can be seen in Figures 2A and B, with the taller domain comprising a network of circular shapes connected by fractal-like strands with widths varying from 50 to 500 nm, similar to those seen in previous studies.[18-19] Additionally, phase imaging of the mixed monolayer film with tapping mode AFM resolved individual nanoparticles with diameters of ~ 6 nm arranged in a hexagonally close packed structure within the taller domains (Figure 2C). Finally, a line segment analysis across the two distinct domains (Figure 2D) reveals a step of ~ 6 nm from the shorter to the taller phase. A 6 nm step between the two domains is consistent with the height difference between a monolayer of DPPC, which has a height of 2 nm (darker region), and a monolayer of Au-NPs, which has a height of ~ 8 nm from the combination of the Au-NP core and dodecanethiol ligands (brighter region). These observations indicate that a phase separation exists between the DPPC and Au-NPs in the mixed monolayer, and that the interaction between the two materials is minimized through clustering of the Au-NPs in large domains and networks. The domains and networks in turn create cages around large regions of DPPC. It should be noted, however, that the resolution of AFM imaging is insufficient to definitively rule out changes in the internal packing of the Au-NP domains through intermixing of small quantities of DPPC molecules into the Au-NP network. Furthermore, although AFM can probe the relative thickness of the two phases, information regarding the topography of the Au-NP and DPPC domains relative to the air/water interface is lost during the transfer of the film from the air/water interface to a solid substrate for AFM imaging.

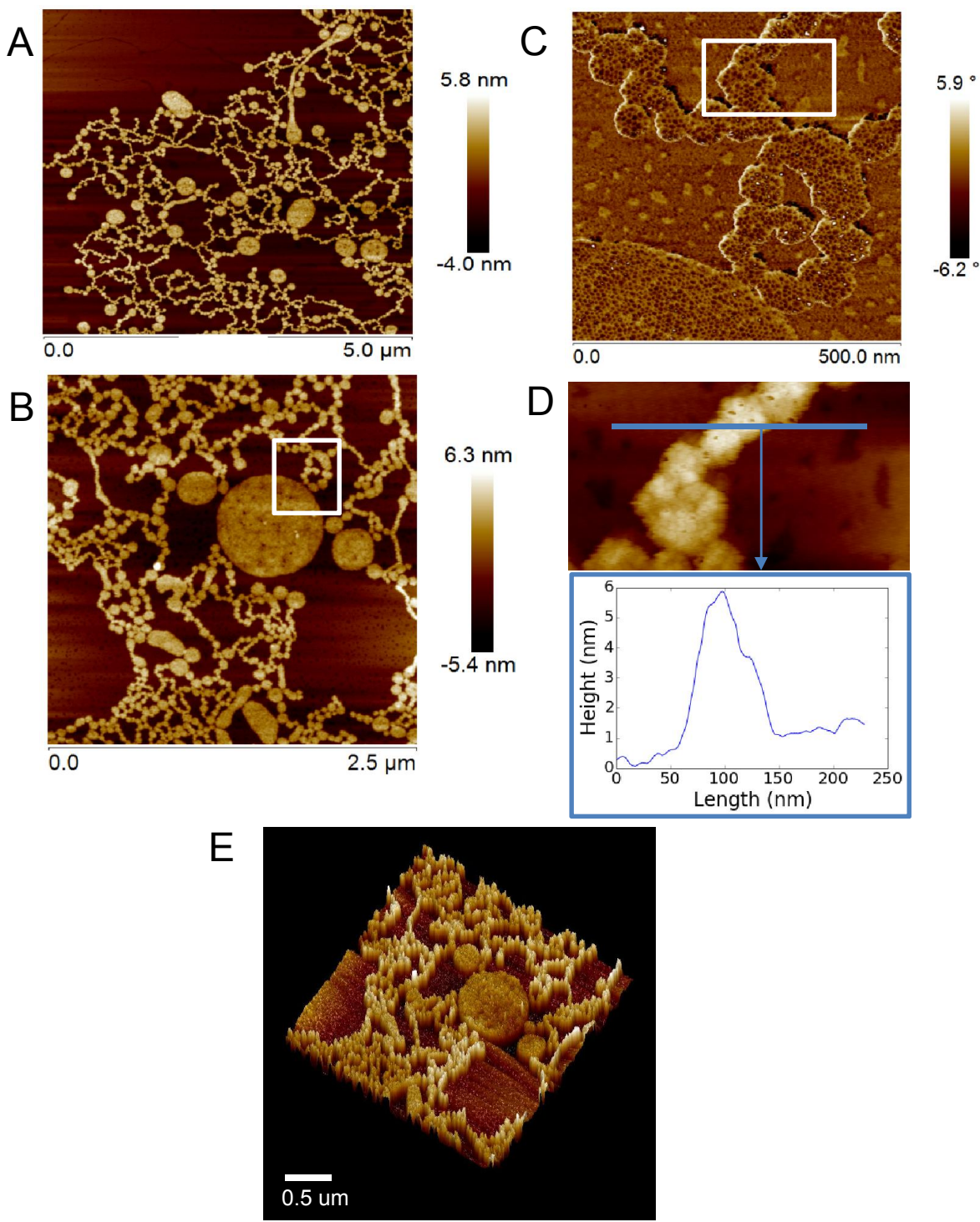

**Figure 2:** AFM images of DPPC/Au-NP films at an Au-NP surface coverage of (A) 5% and (B) 20% imaged in tapping mode. (C) Imaging the portion contained within the white square in panel B in phase mode reveals hexagonally close packed nanoparticle islands. (D) A line profile height analysis (indicated by a blue line) within the boxed region in panel C reveals a height difference of ~ 6 nm, in agreement with the height difference between a monolayer of Au-NPs and a monolayer of DPPC. (E) A 3D image of the structure in panel B.

Liquid surface X-ray scattering techniques were then used to obtain structural information on the composite Au-NP-DPPC system *in situ*. By measuring the XR from the interface as a function of $q_z$ while keeping $q_{xy}$ =0 and fitting the data using the Parratt method,[31, 48-49] the electron density profile of the film near the air/water interface can be measured. A schematic of the DPPC/Au-NP system is shown in Figure 3A.

The XR for pure DPPC and pure Au-NP films were first measured and fit, as shown in Figure 3B. Both data sets exhibit symmetrical Kiessig fringes, an indication of uniform layering.[21] Electron density profiles (EDP) of the pure Au NP and pure DPPC films were extracted from the XR fit and shown in Figure 3C. For a pure DPPC film, there are two distinct regions to the EDP: the hydrophobic tail, which is electron poor relative to the water subphase and faces the air superphase, and the hydrophilic head group, which is electron rich relative to the water subphase and is adjacent to the subphase layer.[52] The total distance of these layers is approximately the length of the extended lipid molecule (~ 2 nm). For the Au-NP film, a broad peak of ~ 6 nm in height is observed corresponding to electron density of the gold core. As the dodecanthiol ligands are relatively poor in electron density, they do not contribute to any significant features in the XR. These measured EDPs agree well with previous results.[29-31, 33]

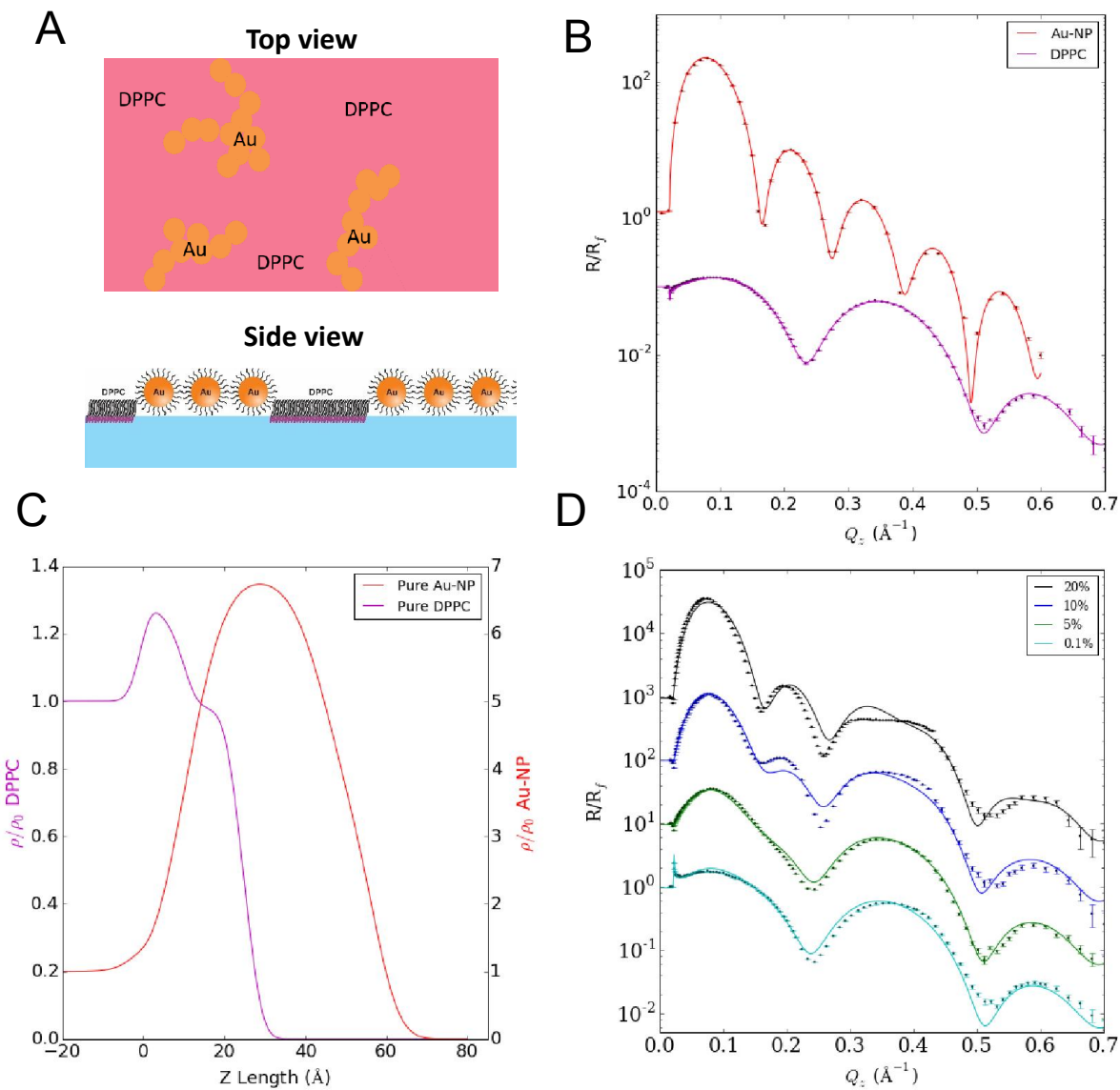

**Figure 3:** **(**A) Schematic representation of DPPC/Au-NP on liquid surface. (B) Reflectivity and Parratt model fit for films of Au-NP and pure DPPC (offset by factor of 10 for ease of viewing). (C) Normalized electron density profiles for pure Au-NP and pure DPPC films calculated from Parratt fitting. (D) XR reflectivity and Parratt model fitting with fixed DPPC and Au-NP reflectivity. Each successive plot is offset by a factor of ten for ease of viewing.

The plots in Figure 3D illustrate the measured XR of the DPPC/Au-NP mixtures with four different Au-NP surface densities, all exhibiting irregular fringes. However, the positions of reflectivity peaks from individual DPPC and Au-NP systems (Figure 3B) persist. Intuitively, the XR of mixtures could be a linear combination of the XR of pure DPPC and pure Au-NP with a proper weighting. This observation, combined with the AFM images showing phased separated domains of DPPC and Au-NP (Figure 2), gives a clear hint that the XR of mixtures can be modeled by incoherent scattering from phase separated domains of DPPC and Au-NP, schematically illustrated in Figure 3A. Therefore, the XR of mixtures could be fitted by the equation $R_{mix} = aR_{DPPC} + (1-a)R_{Au-NP}$ where $a$ is the relative surface area coverage of DPPC, and $R_{DPPC}$ and $R_{Au}$ are the fixed, fitted reflectivity response of the pure DPPC and Au-NP films, respectively shown in Figure 3B. Modeling by incoherent scattering with $a$ as the only fitting parameter yields reasonable fits for all four sets of data with different Au-NP surface densities (Figure 3D). The relative surface area fraction, $a$, extracted from the fitting agrees with the estimated surface area ratio made for the calculation of the Au-NP surface density (see Table 2). The discrepancies between the experimental data and fits may have two origins. First, mixing might alter the individual organizational structure of the separate domains of Au-NP and DPPC. An attempt to fit with varying EDP of DPPC and Au-NP substantially improves the fit (data not shown) though it is not clear if the calculated alterations in the EDP are physically meaningful. Second, domain size of the Au-NP islands as seen from the AFM imaging (Figure 3A) may be smaller than the x-ray coherence length (~10 μm), especially at low Au-NP densities, reducing the accuracy of the incoherent scattering fitting.

| DPPC/Au-Au Calculated Area Ratio | $a$ obtained from fitting with fixed DPPC and Au-NP EDP |
|---|---|
| **99:1** | 99.7 |
| **95:5** | 98.7 |
| **90:10** | 94.7 |
| **80:20** | 85.9 |

**Table 2.** The calculated *a* fitting factor from incoherent scattering assumption for the various XR fitting methods compared to the previously estimated DPPC/Au-NP area coverage.

In addition to reflectivity measurements, GIXD was conducted to probe the structure of the composite film in the plane of the interface. The GIXD features of a pure Au-NP monolayer film and a pure DPPC film at 30 mN/m are shown in Figures 4A and 4D, respectively. The size difference in the lattice spacing of DPPC and Au-NPs results in their respective diffraction peaks appearing in different regions of $q_z$ and $q_{xy}$, with no overlap between the respective signals. The key scattering structures in the DPPC film at 30 mN/m are a Bragg peak at $q_{xy} \approx 1.35$ Å$^{-1}$ and positive $q_z$, due to the degenerate (11) and (1$\bar{1}$) reflections of distorted hexagonal packing, and a second Bragg peak at $q_{xy} \approx 1.47$ Å$^{-1}$ and $q_z = 0$ due to the nondegenerate (02) reflection (Figure 4A). In a pure Au-NP film at 30 mN/m (Figure 4A), there are distinct Bragg peaks with the Miller indices (10), ($\bar{1}$1), (20), and (2$\bar{1}$), corresponding to a hexagonal packing of the nanoparticles in the film. The $q_{xy}$ position of these peaks is inversely proportional to the spacing of the Au-NP. Beyond decreased intensity proportional to the sample Au-NP content, GIXD measurements of composite films (Figures 4B and 4C) revealed some changes to the diffraction signal for Au-NPs relative to pure Au-NP films.

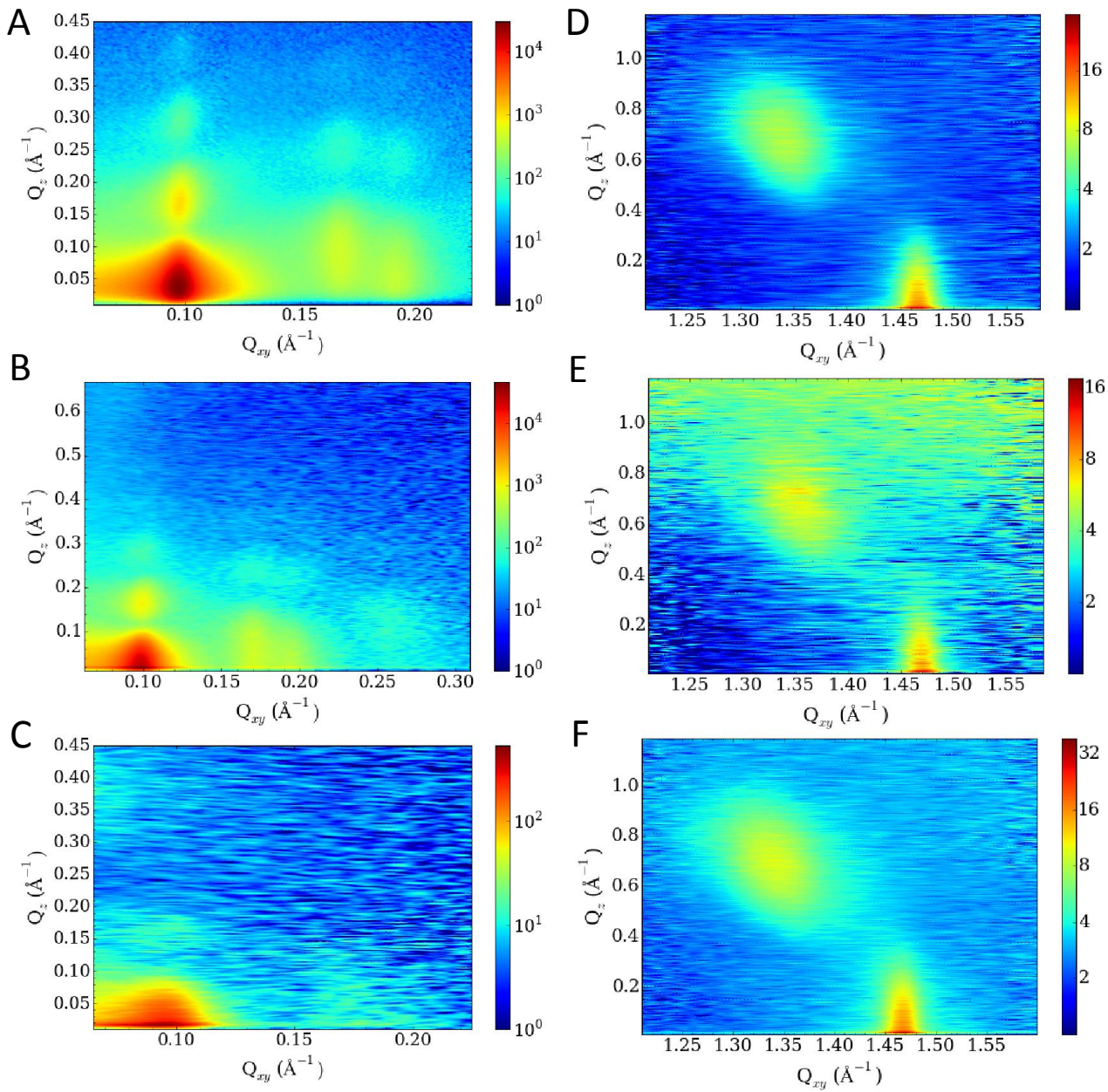

**Figure 4:** GIXD measurements reveal in-plane structure of DPPC/Au-NP composite films at the air/water interface. The Au-NP diffraction pattern is shown for (A) a pure Au-NP film, (B) a composite DPPC/Au-NP film at a Au-NP area coverage of 5%, and(C) a composite DPPC/Au-NP film at a Au-NP area coverage of 0.1%. The DPPC diffraction pattern is shown for (D) a pure DPPC film, (E) a composite DPPC/Au-NP film at a Au-NP area coverage of 5%, and (F) a composite DPPC/Au-NP film with Au-NP coverage of 0.1%. Despite the increased background scattering caused by diffuse scattering from the high electron density of the Au-NPs in the film, it is apparent that the DPPC diffraction peaks are at close to the same $q_{xy}$ and $q_z$ as those found in pure DPPC films.

Notably, there is an increase in the full width half maximum of the 1st order diffraction peak corresponding to a decrease in the correlation length of the Au-NP islands (Figure 5, Table 3).

This observation suggests that the size of the nanoparticle islands in the composite film decreases with decreasing Au-NP surface area ratio. At very low surface area coverage, such as 0.1% Au-NP, there is an insufficient number of Au-NPs to form the fractal-like networks, and the Au-NPs form small monolayer "rafts" as observed in optical imaging, which may explain the noticeably shorter correlation lengths indicated in GIXD measurements of films with low Au-NP content. However, despite the change in the FWHM of the GIXD peak, there are no substantial changes in the Au-NP peak position with decreasing Au-NP coverage. The Au-NP core-core distance, calculated from the first order diffraction peak as $d = \frac{2\pi}{q_{10}} \sin\left(\frac{\pi}{3}\right)$ [30] remains around ~ 77 Å, which is close to the combined length of a single nanoparticle core and a dodecanthiol ligand, assuming the ligands of the Au-NP are fully interdigitated. The variation in the peak position is more indicative of variation or inhomogeneity of the film rather than any effect from the introduction of the DPPC into the film. This indicates that the DPPC component within the composite film does not influence the local ordering of the Au-NP structure, supporting the idea that the two components of the film remain phase-separated.

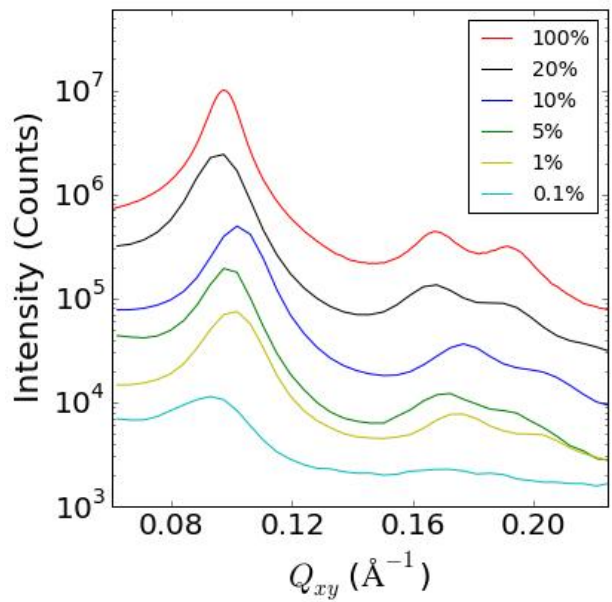

**Figure 5:** Stacked Au-NP diffraction signal integrated over $q_z$ from a pure Au-NP film (red), and composite DPPC/Au-NP films at Au-NP content of 20% (black), 10% (blue), 5% (green), 1% (yellow), and 0.1% (cyan) area coverage. The pure Au-NP film has intensity multiplied by 10 for clarity. Comparison of the Au-NP diffraction signal from the composite films to that of the pure Au-NP film shows that decreasing Au-NP content in the composite films has almost no effect on the peak position. Full width at half maximum of the peak increases with decreasing Au-NP content, suggesting that films with lower Au-NP content have smaller domain size.

| Au-NP Area % | 1st Order Diffraction Peak $Q_z$ | Full Width Half Max (Å$^{-1}$) | Au-NP Center-to-Center (Å) | Coherence Length (Å) |
|---|---|---|---|---|
| 100 | 0.0964 | 0.0115 | 75 | 540 |
| 20 | 0.0976 | 0.0156 | 74 | 402 |
| 10 | 0.102 | 0.0157 | 71 | 400 |
| 5 | 0.0943 | 0.0154 | 77 | 407 |
| 1 | 0.0998 | 0.0168 | 73 | 374 |
| 0.1 | 0.0932 | 0.0242 | 78 | 260 |

**Table 3.** First order Au-NP diffraction peak positions and FWHM obtained from fitting of the first order diffraction peaks upon $z$ integration of the GIXD data obtained for the different composite DPPC/Au-NP films. Au-NP center-to-center distance and coherence length are calculated from these results.

Although mostly unaffected, the diffraction signals of DPPC in the composite films show several minor changes relative to a film of pure DPPC (Figure 4D). As the content of Au-NPs in the composite film is increased, the background scattering in the DPPC diffraction signal, especially at higher values of $q_z$, increases (Figures 4E, 4F) due to the random diffuse scattering from the substantial electron density of the Au-NPs. This results in a decay of the signal-to-noise ratio of the DPPC diffraction pattern with increasing Au-NP content in the composite film. Despite the decaying signal-to-noise ratio as the content of the Au-NPs is increased, a slight shift in the peak position and spread in $q_z$ and $q_{xy}$ for both the out-of-plane and the in-plane peaks are clearly observed. Although small, these shifts are greater than the minimum resolution of our instrumentation. Rather than indicating a complete disruption of the packing structure of the DPPC, the small shifts in the position and spread of the DPPC diffraction peaks (Table 4) indicate a distortion of the unit cell of the DPPC region of the composite film.[26] Due to a size mismatch of the larger head group and smaller tail group, the tails of DPPC are in a tilted orientation with respect to the interface normal when assembled in a monolayer. The tilt angle, $\theta$, for nearest neighbor tilt is given by: [26]

$$tan\,\theta = q_z \left( q_{11}^2 - \left(\frac{q_{02}}{2}\right)^2 \right)^{-\frac{1}{2}}$$

As the Au-NP content in the composite film is increased, the shift in the DPPC diffraction peaks reveals a decrease in the angle of the DPPC tails with respect to the interface normal (Figure 6). A decrease in the tilt angle of the DPPC tails could be caused by the rigidity of the Au-NP spheres that surround domains of lipids and restrict the lipid tail groups from adopting their preferred tilt angle.[23]

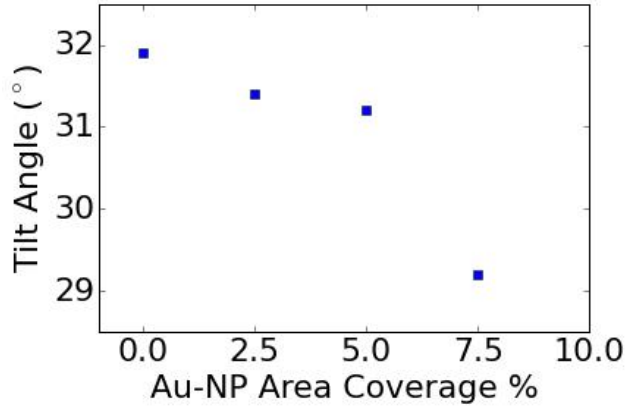

**Figure 6:** Tilt angle of the DPPC tail group with respect to the air/water interface normal as a function of the Au-NP content in DPPC/Au-NP composite films. For composite films with Au-NP area coverage greater than 7.5 %, the signal to noise ratio of the DPPC diffraction peaks is too weak to reliably fit for the extraction of the lipid tail tilt angle.

| **Au-NP Calculated Area Coverage** | $q_{02}$ (Å⁻¹) | $q_{11}$ (Å⁻¹) | $q_z$ (Å⁻¹) | $\Theta(°)$ |
|---|---|---|---|---|
| **0** | 1.466 | 1.355 | 0.710 | 31.9 |
| **2.5 %** | 1.468 | 1.354 | 0.694 | 31.4 |
| **5 %** | 1.469 | 1.355 | 0.689 | 31.2 |
| **7.5 %** | 1.470 | 1.362 | 0.640 | 29.2 |

**Table 4.** Reciprocal space lattice parameters obtained from GIXD data for composite films of Au-NPs and DPPC as a function of Au-NP content: $q_{02}$ and $q_{11}$ are the lattice parameters in Fourier space for the two strong reflections from the $q_z$-averaged data; $q_z$ is the z-coordinate of the (11) reflection. All (02) reflections are centered at $q_z = 0$. $\theta$ is defined as the tilt angle of the lipid tail from the normal of the liquid surface.

## Conclusion

Langmuir monolayers of composite films with low Au-NP content (up to 2.5 %) retain the broad mechanical properties of pure DPPC films. The stress response, as measured with surface pressure-surface area isotherms, is dominated by the DPPC molecules present. As seen with XR measurements, the two components of the composite film remain phase separated at a surface pressure of up to 30 mN/m.

However, as the Au-NP content in the composite film increases, the structural organization of the film changes. At low Au-NP content, the composite film is best characterized as a film of DPPC containing isolated islands of Au-NPs, judging from the lack of higher order Bragg scattering peaks. The density of the islands of Au-NPs increases with increasing Au-NP area coverage up to ~ 5 %, where the intensity of the Au scattering background increases substantially, and higher order Au-NP Bragg scattering peaks can be observed. AFM reveals that the Au-NPs form an interconnected fractal-like structure at this point.

Confirming this change in the structural makeup of the composite film beyond an Au-NP average coverage of 5% is a marked shift in the tilt of the DPPC tail group with respect to the interface normal as measured with GIXD. As the DPPC domains become fenced in by the increasing presence of Au-NPs, the large Au-NPs set a hard boundary at the edge of the DPPC domains, restricting the freedom of their hydrophobic tails to attain a more tilted state as seen in the absence of Au-NPs.

Our findings show the potential for nanomaterials to effect change in the structure and ordering of lipid membranes, and demonstrate how LSXS techniques can be useful in monitoring and assessing these nano-bio interactions *in situ*.

**Author Information:**

Corresponding Authors:

*Email: lin@cars.uchicago.edu (B. L.)

*Email: kayeelee@uchicago.edu (K.Y.C.L)

**Acknowledgments:**

We thank L. Boucheron, J. Stanley and S. Griesemer for their experimental help and insightful discussions, and Prof. S. A. Rice for the lending of equipment and fruitful discussions. This work was supported by the National Science Foundation through the MRSEC program at the University of Chicago (DMR-1420709). J.M.H. acknowledges partial support from the NSF (MCB-1413613). ChemMatCARS Sector 15 is supported by the National Science Foundation under Grant No. NSF/CHE-1346572. This research used resources of the Advanced Photon Source, a U.S. Department of Energy (DOE) Office of Science User Facility operated for the DOE Office of Science by Argonne National Laboratory under Contract No. DE-AC02-06CH11357.

**Table of Contents Only, TOC Figure**

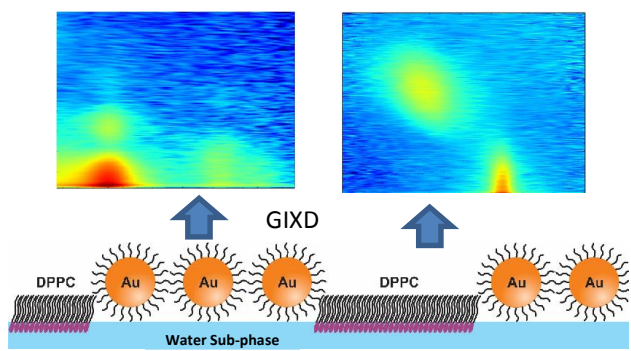